\PassOptionsToPackage{unicode}{hyperref}
\PassOptionsToPackage{hyphens}{url}
\documentclass[
]{article}
\usepackage{xcolor}
\usepackage{amsmath,amssymb}
\usepackage{iftex}
\ifPDFTeX
  \usepackage[T1]{fontenc}
  \usepackage[utf8]{inputenc}
  \usepackage{textcomp} 
\else 
  \usepackage{unicode-math} 
  \defaultfontfeatures{Scale=MatchLowercase}
  \defaultfontfeatures[\rmfamily]{Ligatures=TeX,Scale=1}
\fi
\usepackage{lmodern}
\ifPDFTeX\else
\fi
\IfFileExists{upquote.sty}{\usepackage{upquote}}{}
\IfFileExists{microtype.sty}{
  \usepackage[]{microtype}
  \UseMicrotypeSet[protrusion]{basicmath} 
}{}
\makeatletter
\@ifundefined{KOMAClassName}{
  \IfFileExists{parskip.sty}{%
    \usepackage{parskip}
  }{
    \setlength{\parindent}{0pt}
    \setlength{\parskip}{6pt plus 2pt minus 1pt}}
}{
  \KOMAoptions{parskip=half}}
\makeatother
\usepackage{longtable,booktabs,array}
\usepackage{calc} 
\usepackage{etoolbox}
\makeatletter
\patchcmd\longtable{\par}{\if@noskipsec\mbox{}\fi\par}{}{}
\makeatother
\IfFileExists{footnotehyper.sty}{\usepackage{footnotehyper}}{\usepackage{footnote}}
\makesavenoteenv{longtable}
\usepackage{graphicx}
\makeatletter
\newsavebox\pandoc@box
\newcommand*\pandocbounded[1]{
  \sbox\pandoc@box{#1}%
  \Gscale@div\@tempa{\textheight}{\dimexpr\ht\pandoc@box+\dp\pandoc@box\relax}%
  \Gscale@div\@tempb{\linewidth}{\wd\pandoc@box}%
  \ifdim\@tempb\p@<\@tempa\p@\let\@tempa\@tempb\fi
  \ifdim\@tempa\p@<\p@\scalebox{\@tempa}{\usebox\pandoc@box}%
  \else\usebox{\pandoc@box}%
  \fi%
}
\def\fps@figure{htbp}
\makeatother
\usepackage{bookmark}
\IfFileExists{xurl.sty}{\usepackage{xurl}}{} 
\hypersetup{
  hidelinks,
  pdfcreator={LaTeX via pandoc}}

\author{}
\date{}

\begin{document}

\footnote{}

A Logistic Regression Model to Predict Malaria Severity in Children

Mary O. Ansong, Asare Y. Obeng and Samuel K. Opoku

\emph{Abstract} --- One of the main causes of death around the globe is
malaria. Researchers have sought to develop predicting models for
malaria outbreaks based on metrological data, climate data and the
breeding cycle of plasmodium, the causative agent of malaria. This study
predicts the severity of malaria based on environmental and biological
factors. A logistic regression model was developed in this study to
predict the severity of malaria based on such factors as sickle cell
disease, stagnant water, garbage dump, wet lawns, and the use of treated
mosquito nets with an 83.3\% accuracy rate. The study was carried out in
the Bosomtwe District of Ghana with 417 respondents. It was deduced that
although children in the District are highly prone to malaria infection,
the severity is very low. The study recommends that not just having a
good sample size alone is important during machine learning model
development but also having a good sample representation of the various
class labels is equally important.


\section{Introduction}\label{introduction}

The causative agent of Malaria is a protozoan parasite called
Plasmodium. The deadly type of the Plasmodium parasite is the
\emph{Plasmodium falciparum} causing about 90\% fatalities of malaria in
humans in Africa. Malaria has become a public health issue in Africa
with the fatality rate increasing exponentially in children under five
years old {[}1{]}, {[}2{]}. In 2019, half of the global cases of malaria
amounting to about 229 million came from six countries in Africa with
Nigeria contributing to 25\% of the global count {[}3{]}. The initial
clinical symptom, usually fever, is followed by vomiting, tiredness,
abdominal pain and diarrhoea. Failure to treat falciparum malaria within
24 hours after observing the initial clinical symptom can be fatal.
Symptoms can be complicated to the extent of organ or system failure.
One of the common systems usually attacked by malaria is the central
nervous system resulting in cerebral malaria {[}3{]}, {[}4{]}.

Malaria is transmitted to humans by infected female Anopheles mosquitoes
as the male Anopheles mosquitoes do not suck blood {[}5{]}. Malaria can
also be transmitted from a pregnant mother to her baby, through blood
transfusion and sharing of needles used to inject drugs {[}6{]}. Fig. 1
illustrates the transmission cycle of malaria

\includegraphics[width=3.38955in,height=2.37199in]{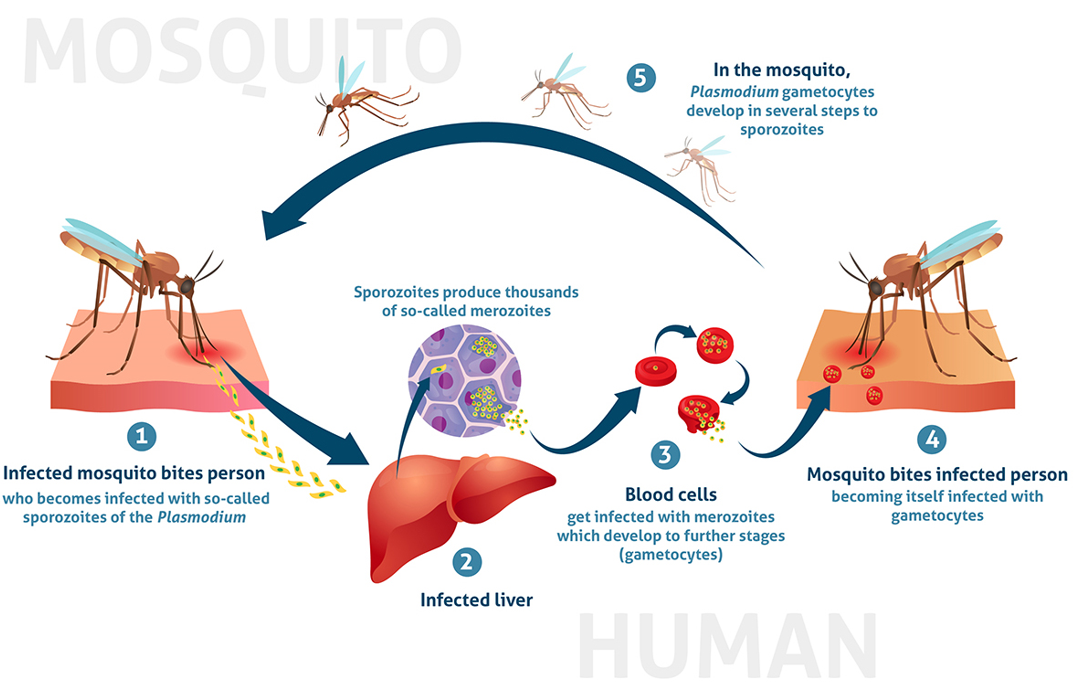}

Fig. 1. Transmission Cycle of Malaria Parasite (Source {[}3{]})

From Fig. 1, the upper section, labelled 5, shows the development of the
plasmodium parasite from gametocytes to sporozoites which can affect the
human host in mosquitoes. The lower section, labelled 1 to 4,
illustrates how the human host is affected. An infected mosquito bites
the host and transmits the sporozoites of the Plasmodium, the malaria
parasite, into the host. The parasite travels to the liver of the host
where it lies dormant for about ten to twenty-eight days. In the liver,
the sporozoites develop thousands of merozoites. The merozoites leave
the liver and infect red blood cells. This is when malaria signs and
symptoms develop. The merozoites develop into further stages called
gametocytes. A mosquito bites an infected person and it is infected with
gametocytes. The transmission cycle continues from point 5 in Fig. 1.

There have been various studies on malaria prediction {[}7{]}-{[}11{]}.
However, the techniques of data collection and analysis of the illness
prevalence and breadth varied greatly. Some of the methods employed in
malaria predictions are Naive Bayes {[}7{]}, {[}12{]}, Support Vector
Machine (SVM) {[}7{]}, {[}12{]}, Linear Regression {[}7{]}, Logistic
Regression {[}7{]}, K-Nearest Neighbor (KNN) {[}7{]}, {[}12, Random
Forests {[}12{]} deep learning {[}13{]}, multi-layer perceptron and
decision tree, {[}10{]}. Unfortunately, predicting malaria outbreaks
using linear regression has not been successful {[}7{]}. None of the
existing works has predicted the severity of malaria in children.

Researchers train their machine-learning models using data with various
attributes. Some of the researchers predicted malaria outbreaks using
meteorological and malaria incident data {[}7{]}, meteorological data
and the malaria-carrying vector (mosquito) breeding environment
{[}13{]}, climatic conditions {[}10{]}, biological characteristics and
social determinants associated with demographic and health survey
{[}12{]}. None of the research works mentioned above predicted the
severity of malaria based on the combination of environmental and
biological factors.

This study used a logistic regression model to predict the level of
malaria severity in children using environmental conditions and
biological factors. Different methodologies have their own set of
advantages and disadvantages. Logistic regression facilitates the
conversion of complex data into simple and relevant insights. By
utilizing algorithms to answer queries and have conversations, logistic
regression aims to imitate human-like responses {[}14{]}. Logistic
regression can be used in the healthcare sector to summarize large
narrative texts, such as academic journal articles or clinical notes, by
highlighting essential concepts or phrases in the reference document. To
improve clinical decision-making, logistic regression can map data
pieces in Electronic Health Records that are available as unstructured
text into structured useful data {[}15{]}. The remaining part of this
paper is divided into three sections that focus respectively on the
materials and methods, results and conclusion.

\section{Materials and Methods}\label{materials-and-methods}

The model to predict the severity of malaria in children was developed
using a logistic regression algorithm. This algorithm requires that
certain conditions are satisfied to develop a good model. The logistic
regression algorithm ensures that the dependent variable is binary with
the value 1 representing the desired outcome. Certain independent
variables which are not meaningful and depend on other independent
variables (collinearity) should be avoided. The categorical data should
be numerically coded. A serious limitation of the logistic regression
algorithm is associated with overfitting where the model learns both the
data and the associated noise so well that it is not able to perform
well with unseen data. Logistic regression requires an appreciable size
of data for training. This section describes the various methods that
were carried out to develop a good model. It also explains how the
conditions required by the logistic regression algorithm were handled.

\subsection{\texorpdfstring{Data Collection and Data Processing
}{Data Collection and Data Processing }}\label{data-collection-and-data-processing}

The study population was narrowed down to Amakom in the Bosomtwe
District of Ghana to determine the severity of malaria in children under
five years due to the high occurrence of malaria in the area {[}16{]}.
The study population is two thousand. A sample size of four hundred and
seventeen (417) children using the Stratified sampling method was ideal
for the study {[}17{]}. The dataset was examined, cleaned and processed
for analysis and machine learning model development.

\subsection{Problem Formulation}\label{problem-formulation}

Given the features or attributes of the independent variables
\(F = (f_{i},{\ f}_{2},\ f_{3}\ .\ .\ .\ ,\ f_{T})\) where T is the
number of features or attributes and the corresponding severity of
malaria \(\ S \in \{ 0,\ 1\}\) where 0 means no severe malaria and 1
means severe malaria. We determine a logistic regression function LR (f)
such that the predicted output LR (f\textsubscript{i}) are possibly
close to the actual output S\textsubscript{i} for each observation
\(i = 1,\ 2,\ \ldots,\ n\ \) where n is the number of observation. Given
that\({\ S}_{i} \in \left\{ 0,\ 1 \right\}\ \forall LR\ \left( f_{i} \right) \approx 0\ |1\),
the sigmoid function \(\sigma(x) = \ \frac{1}{1 + \ \exp^{( - x)}}\) was
used with the logistic regression function.

Since a logistic function is a linear classifier, it implies

\[f(x) = \ \sum_{j = 0}^{T}{b_{j}x_{j}\ such\ that\ \forall x_{0}\  = 1}\]

The\({\ b}_{j}'s\ such\ that\ \ \forall j = 0,\ 1,\ 2,\ \ldots,\ T\) are
the predicted weights which are the estimated coefficients. The logistic
regression function LR (f) is a sigmoid function given as:

\[f(x)\ :\ LR(f) = \frac{1}{1 + \ \exp^{( - f(x))}}\  \approx 0\ |1\ \]

The predicted weights
(\({\ b}_{j}\ \forall j = 0,\ 1,\ 2,\ \ldots,\ T\)) are best obtained by
maximizing the log-likelihood function (LF) for all the observations as

\[\ LF = \ \ \sum_{i}^{}{{(S}_{i}\log\left( LR\left( f_{i} \right) \right) + \left( 1 - \ S_{i} \right)\log{(1 - LR(f_{i})))}}\]

The calculation of the best weights using the LF was handled by the
logistic regression Python libraries. There are such open-source
packages in Python as NumPy for manipulating arrays and Matplotlib to
visualize results. The scikit-learn and statsModels are the libraries
used to create, fit (or train), evaluate (or test) and apply a logistic
regression model. To avoid multicollinearity, the correlation between
the independent variables was determined.

\subsection{Handling Overfitting of Logistic Regression
Algorithm}\label{handling-overfitting-of-logistic-regression-algorithm}

One way to control overfitting is by regularizing the predicted weights.
There are three methods of penalizing large coefficients in logistic
regression to control overfitting. The first method, usually called L1,
uses the absolute values of the predicted weights such that the linear
function becomes:

\[f(x) = \ \sum_{j = 0}^{T}{{|b}_{j}|x_{j\ }\ such\ that\ \forall x_{0}\  = 1}\]

Another method, usually called L2, uses the squared values of the
predicted weights so that the linear function becomes:

\[f(x) = \ \sum_{j = 0}^{T}{{b_{j}}^{2}x_{j}\ such\ that\ \forall x_{0}\  = 1}\]

The final method, usually called elastic-net, combines the previous two
methods and the linear function becomes:

\[f(x) = \ \sum_{j = 0}^{T}{{{|b}_{j}|}^{2}x_{j}\ such\ that\ \forall x_{0}\  = 1}\]

Scikit-learn library allows regularization setting using the penalty
parameter. The scikit-learn library was therefore chosen over
statsModels. The scikit-learn library was used with the logistic
regression class implementing the `liblinear' library which requires
regularization to work using L1, L2 or elasticnet as values to the
penalty parameter {[}18{]}. The default penalty is L2 which was used as
demonstrated in some research works {[}19{]}, {[}20{]}.

\section{Results}\label{results}

This section presents the analysis of data collected from four hundred
and seventeen (417) self-administered questionnaires. These were parents
who completed the questionnaire on behalf of their children under five
years. The findings were presented in tables and figures to complement
the interpretation of the data collected.

\subsection{Descriptive Statistics of the Data
Collected}\label{descriptive-statistics-of-the-data-collected}

This section describes the features of the data by generating summaries
of the data collected. Table I describes how often respondents' children
get malaria every year.

TABLE I: How Often Children Got Malaria

{\def\LTcaptype{none} 
\begin{longtable}[]{@{}
  >{\centering\arraybackslash}p{(\linewidth - 4\tabcolsep) * \real{0.1416}}
  >{\centering\arraybackslash}p{(\linewidth - 4\tabcolsep) * \real{0.0991}}
  >{\centering\arraybackslash}p{(\linewidth - 4\tabcolsep) * \real{0.0991}}@{}}
\toprule\noalign{}
\endhead
\bottomrule\noalign{}
\endlastfoot
Often Malaria & Frequency & Percentage \\
No Malaria & 204 & 48.920863 \\
Up to 2 times & 157 & 37.649880 \\
3 to 5 times & 33 & 7.913669 \\
More than 5 times & 23 & 5.515588 \\
Total & 417 & 100 \\
\end{longtable}
}

Table I shows that 204 (48.9\%) of the respondents\textquotesingle{}
children often do not get malaria every year, though the respondents'
children had contracted malaria before but it does not occur every year.
213 (51.1\%) of the respondents\textquotesingle{} children contract
malaria from one to more than five times a year. This shows how
prevalent malaria occurs in the area.

Table II shows the severity of malaria in respondents' children.
Respondents were asked to describe the severity of malaria that affect
their children

TABLE II: Severity of Malaria

{\def\LTcaptype{none} 
\begin{longtable}[]{@{}
  >{\centering\arraybackslash}p{(\linewidth - 4\tabcolsep) * \real{0.1416}}
  >{\centering\arraybackslash}p{(\linewidth - 4\tabcolsep) * \real{0.0991}}
  >{\centering\arraybackslash}p{(\linewidth - 4\tabcolsep) * \real{0.0991}}@{}}
\toprule\noalign{}
\endhead
\bottomrule\noalign{}
\endlastfoot
Severity & Frequency & Percentage \\
Not severe & 336 & 80.58 \\
Severe & 48 & 11.51 \\
Very severe & 18 & 4.32 \\
Resulted in death & 15 & 3.60 \\
Total & 417 & 100.0 \\
\end{longtable}
}

Table II shows that 336 (80.6\%) of malaria occurrence is not severe.
These children do not stay in clinics or hospitals for more than 24
hours. However, 81 (19.4\%) of malaria occurrence is severe or fatal.
These children usually stay in hospitals for more than 24 hours and they
are placed in intensive care units.

Table III summarises the environmental conditions of the respondents. It
shows the number of respondents answering Yes and those answering No for
the selected environmental conditions.

TABLE III: Environmental Conditions

{\def\LTcaptype{none} 
\begin{longtable}[]{@{}
  >{\centering\arraybackslash}p{(\linewidth - 10\tabcolsep) * \real{0.0765}}
  >{\centering\arraybackslash}p{(\linewidth - 10\tabcolsep) * \real{0.0852}}
  >{\centering\arraybackslash}p{(\linewidth - 10\tabcolsep) * \real{0.0851}}
  >{\centering\arraybackslash}p{(\linewidth - 10\tabcolsep) * \real{0.0852}}
  >{\centering\arraybackslash}p{(\linewidth - 10\tabcolsep) * \real{0.0818}}
  >{\centering\arraybackslash}p{(\linewidth - 10\tabcolsep) * \real{0.1136}}@{}}
\toprule\noalign{}
\endhead
\bottomrule\noalign{}
\endlastfoot
Answer & A lot of Mosquito & Sleep under Mosquito net & Wet Lawns in the
residence & Refuse Dump within 300m & Constant Stagnant Water around the
residence \\
Yes & 261 & 250 & 218 & 208 & 267 \\
No & 156 & 167 & 199 & 209 & 150 \\
Total & 417 & 417 & 417 & 417 & 417 \\
\end{longtable}
}

From Table III, the environmental conditions considered in this study
were the presence of mosquitoes in the area. Respondents answered this
question based on mosquito sounds they hear and whether they see them
physically. Other environmental conditions looked at the nature of the
environment, namely, the occurrence of wet lawns in the respondents'
residence or close to the residence; refuse dump within 300 meters of
the respondents' residence and constant stagnant water in or around the
respondents' residence.

A biological factor which focuses on the genotype of sickle cell Anemia
in respondent children was described in Table IV

TABLE IV: Sickle Cell Anemia Genotype of Children

{\def\LTcaptype{none} 
\begin{longtable}[]{@{}
  >{\centering\arraybackslash}p{(\linewidth - 4\tabcolsep) * \real{0.1531}}
  >{\centering\arraybackslash}p{(\linewidth - 4\tabcolsep) * \real{0.0992}}
  >{\centering\arraybackslash}p{(\linewidth - 4\tabcolsep) * \real{0.1132}}@{}}
\toprule\noalign{}
\endhead
\bottomrule\noalign{}
\endlastfoot
Sickle Cell Anemia & Frequency & Percentage \\
Yes & 203 & 48.68 \\
No & 214 & 51.32 \\
Total & 417 & 100 \\
\end{longtable}
}

Table IV shows that 203 (48.68\%) were sickle cell anaemic or carriers.
These children have S-gene as either SS or AS genotypes. However, 214
(51.32\%) were considered non-sickle cell anaemic. They do not have
S-gene, though they may have the C-gene. They are considered AA or AC
genotypes.

\subsection{Collinearity of Independent
Variables}\label{collinearity-of-independent-variables}

To ensure that the independent variables do not depend on themselves and
avoid multicollinearity, a correlation matrix between the independent
variable was calculated. Table V shows the correlation results.

TABLE V: Correlation Results

{\def\LTcaptype{none} 
\begin{longtable}[]{@{}
  >{\centering\arraybackslash}p{(\linewidth - 14\tabcolsep) * \real{0.1624}}
  >{\centering\arraybackslash}p{(\linewidth - 14\tabcolsep) * \real{0.0511}}
  >{\centering\arraybackslash}p{(\linewidth - 14\tabcolsep) * \real{0.0485}}
  >{\centering\arraybackslash}p{(\linewidth - 14\tabcolsep) * \real{0.0485}}
  >{\centering\arraybackslash}p{(\linewidth - 14\tabcolsep) * \real{0.0453}}
  >{\centering\arraybackslash}p{(\linewidth - 14\tabcolsep) * \real{0.0543}}
  >{\centering\arraybackslash}p{(\linewidth - 14\tabcolsep) * \real{0.0426}}
  >{\centering\arraybackslash}p{(\linewidth - 14\tabcolsep) * \real{0.0427}}@{}}
\toprule\noalign{}
\endhead
\bottomrule\noalign{}
\endlastfoot
& SC & OM & SW & RD & WL & M & SN \\
Sickle cell (SC) & & 0.1 & 0.3 & 0.2 & 0.2 & 0.1 & 0.1 \\
Often malaria (OM) & 0.1 & & 0.1 & 0.1 & 0.1 & 0.0 & 0.1 \\
Stagnant water (SW) & 0.3 & 0.1 & & 0.4 & 0.3 & 0.3 & 0.3 \\
Refuse dump (RD) & 0.2 & 0.1 & 0.4 & & 0.3 & 0.3 & 0.3 \\
Wet lawns (WL) & 0.2 & 0.1 & 0.3 & 0.3 & & 0.4 & 0.3 \\
Mosquitoes (M) & 0.1 & 0.1 & 0.3 & 0.3 & 0.4 & & 0.2 \\
Sleep in net (SN) & 0.1 & 0.1 & 0.3 & 0.3 & 0.3 & 0.2 & \\
\end{longtable}
}

From Table V, the results of correlation among the factors do not have
any effect on each other variables. None of the correlation scores are
above 0.5. This suggests that the variables will serve to give a
meaningful explanation of the model to be developed in this study.

\subsection{Logistic Regression
Results}\label{logistic-regression-results}

Out of 417 records obtained, 333 records were used for training and 84
for testing. Table VI shows the classification report when the model was
tested with the test data.

TABLE VI: Classification Report of Test Data

{\def\LTcaptype{none} 
\begin{longtable}[]{@{}
  >{\centering\arraybackslash}p{(\linewidth - 8\tabcolsep) * \real{0.1480}}
  >{\centering\arraybackslash}p{(\linewidth - 8\tabcolsep) * \real{0.0851}}
  >{\centering\arraybackslash}p{(\linewidth - 8\tabcolsep) * \real{0.0708}}
  >{\centering\arraybackslash}p{(\linewidth - 8\tabcolsep) * \real{0.0851}}
  >{\centering\arraybackslash}p{(\linewidth - 8\tabcolsep) * \real{0.0850}}@{}}
\toprule\noalign{}
\endhead
\bottomrule\noalign{}
\endlastfoot
& Precision & Recall & F1-Score & Support \\
Decrease severity of malaria (No) & 0.88 & 0.93 & 0.90 & 70 \\
Increase severity of malaria (Yes) & 0.50 & 0.36 & 0.42 & 14 \\
Accuracy & & & 0.83 & 84 \\
\end{longtable}
}

From Table VI, the precision examined each side individually,
identifying 0.88 (88\%) No correct on the No side and 0.50 (50\%) Yes
correct on the Yes side. The Recall determined that 0.36 (36\%) correct
Yes and 0.93 (93\%) correct No were on either side of the Yes/No. The
support is identical to the confusion matrix depicted in Table VII.

TABLE VII: Confusion Matrix of the TEST Data

{\def\LTcaptype{none} 
\begin{longtable}[]{@{}
  >{\centering\arraybackslash}p{(\linewidth - 4\tabcolsep) * \real{0.1784}}
  >{\centering\arraybackslash}p{(\linewidth - 4\tabcolsep) * \real{0.1275}}
  >{\centering\arraybackslash}p{(\linewidth - 4\tabcolsep) * \real{0.1630}}@{}}
\toprule\noalign{}
\endhead
\bottomrule\noalign{}
\endlastfoot
& Actual severity & Actual non-severity \\
Predicted severity & 5 & 9 \\
Predicted non-severity & 5 & 65 \\
\end{longtable}
}

Table VII summarizes the performance of the model on the test data,
based on the confusion matrix. Five (5) respondents said their children
had severe malaria and the prediction was the same (true positive).
Sixty-five (65) respondents said their children did not have severe
malaria and the prediction was correct (true negative). Five (5)
respondents said their children have severe malaria but the algorithm
says no (false negative). Nine (9) respondents said their children do
not have severe malaria but the algorithm says yes (false positive). The
outcome shows that we made 70 (65 + 5) correct predictions. According to
the outcome, we made 14 (9+5) incorrect predictions. The accuracy rate
is therefore 83\%.

\section{Conclusion}\label{conclusion}

A logistic regression model was developed in this study to predict the
severity of malaria in children under five years old in Amakom of
Bosomtwe District in Ghana using parameters from environmental and
biological backgrounds with an accuracy rate of 83\%. The model performs
poorly when predicting 1 (that is yes) to indicate the severity of
malaria with 42\% accuracy taking into consideration false positives and
false negatives. This was due to the low representation of data on
malaria severity. However, the model performs creditably well when
predicting 0 (that is no) to indicate non-severity of malaria with 90\%
accuracy taking into account false positives and false negatives. It can
therefore be deduced that although children in Amakom are highly prone
to malaria infection, the severity is very low. Moreover, not just
having a good sample size alone is important during machine learning
model development but also having a good sample representation of the
various class labels is equally important. Future work should focus on
extending the work to cover the whole country to have an appreciable
number of respondents for 1 (that is yes) to indicate the severity of
malaria.

\section{\texorpdfstring{ }{ }}\label{section}

\textbf{References}

{\def\LTcaptype{none} 
\begin{longtable}[]{@{}
  >{\raggedright\arraybackslash}p{(\linewidth - 2\tabcolsep) * \real{0.0402}}
  >{\raggedright\arraybackslash}p{(\linewidth - 2\tabcolsep) * \real{0.4549}}@{}}
\toprule\noalign{}
\begin{minipage}[b]{\linewidth}\raggedright
{[}1{]}
\end{minipage} & \begin{minipage}[b]{\linewidth}\raggedright
Baba E, Hamade P, Kivumbi H, Marasciulo M, Maxwell K, Moroso D, Milligan
P. Effectiveness of seasonal malaria chemoprevention at scale in west
and central Africa: an observational study. \emph{The Lancet.}, 2020;
1829--1840.
\end{minipage} \\
\midrule\noalign{}
\endhead
\bottomrule\noalign{}
\endlastfoot
{[}2{]} & Xu TL, Sun YW, Feng XY, Zhou XN, Zheng B. Development of
miRNA-Based Approaches to Explore the Interruption of Mosquito-Borne
Disease Transmission. \emph{Frontiers in Cellular and Infection
Microbiology.} 2021; 11:665444 \\
{[}3{]} & Biogents.com. \emph{Malaria} {[}Internet{]} 2021 {[}August 16;
cited 2023 August 13{]} Available from:
https://eu.biogents.com/malaria/ \\
{[}4{]} & Hassan AO, Oso OV, Obeagu EI, Adeyemo AT. Malaria Vaccine:
Prospects and Challenges. \emph{Madonna University Journal of Medicine
and Health Sciences.} 2022; 2(2): 22-40. \\
{[}5{]} & Haraguchi A, Takano M, Hakozaki J, Nakayama K, Nakamura S,
Yoshikawa Y, Ikadai H. Molecular mechanisms of Plasmodium development in
male and female Anopheles mosquitoes. \emph{bioRxiv.} 2022; 2022-01. \\
{[}6{]} & Niederhauser C, Galel SA. Transfusion-transmitted malaria and
mitigation strategies in nonendemic regions. \emph{Transfusion medicine
and hemotherapy.} 2022; 49(4): 205-217. \\
{[}7{]} & Stephen A, Akomolafe PO, Ogundoyin KI. A model for predicting
malaria outbreak using machine learning technique. \emph{Annals.
Computer Science Series.} 2020; 9(1):9-15 \\
{[}8{]} & Delgado-Ortet M, Molina A, Alférez S, Rodellar J, Merino A. A
Deep Learning Approach for Segmentation of Red Blood Cell Images and
Malaria Detection. \emph{Entropy.} 2020; 22(6):657 \\
{[}9{]} & Odhiambo JN, Kalinda C, Macharia PM, Snow RW, Sartorius B.
Spatial and spatio-temporal methods for mapping malaria risk: a
systematic review. \emph{BMJ Global Health.} 2020; 5(10):e002919 \\
{[}10{]} & Mohapatra P, Tripathi NK, Pal I, Shrestha S. Determining
suitable machine learning classifier technique for prediction of malaria
incidents attributed to climate of Odisha. \emph{International Journal
of Environmental Health Research}.~2021; \emph{32}(8):1716-1732. \\
{[}11{]} & Yadav SS, Kadam VJ, Jadhav SM, Jagtap S, Pathak PR.~ Machine
learning based malaria prediction using clinical findings.
\emph{International Conference on Emerging Smart Computing and
Informatics.} pp. 216-222, March 2021. \\
{[}12{]} & Sow B, Suguri H, Mukhtar H, Ahmad HF. Using Biological
Variables and Social Determinants to Predict Malaria and Anemia among
Children in Senegal. \emph{IEICE Technical Report; IEICE Tech.
Report.}~2017; 117(336):3-20. \\
{[}13{]} & Masinde M. Africa's Malaria Epidemic Predictor: Application
of Machine Learning on Malaria Incidence and Climate Data. \emph{ACM
International Conference Proceeding Series.} pp. 29-37, 2020. \\
{[}14{]} & Juhn YH. Artificial intelligence approaches using natural
language processing to advance EHR-based clinical research.
\emph{Journal of Allergy and Clinical Immunology.} 2020;
145(2):463-469. \\
{[}15{]} & Bucher BT, Shi J, Ferraro JP, Skarda DE, Samore MH, Hurdle
JF, Finlayson SR. Portable Automated Surveillance of Surgical Site
Infections Using Natural Language Processing: Development and
Validation. \emph{Annals of Surgery,} 2020; \emph{272}(4):629. \\
{[}16{]} & Oteng G, Kenu E, Bandoh D, Nortey P, Afari E.~Compliance with
the who strategy of test, treat and track for malaria control at
Bosomtwi district in Ghana. \emph{Ghana Medical Journal.} 2020;
54(2):40-44. \\
{[}17{]} & Adebayo TS, Odugbesan JA. Modeling CO\textsubscript{2}
emissions in South Africa: empirical evidence from ARDL based bounds and
wavelet coherence techniques. \emph{Environmental Science and Pollution
Research.} 202; 28(8):9377-9389. \\
{[}18{]} & Pedregosa F, Varoquaux G, Gramfort A, Michel V, Thirion B,
Grisel O, Duchesnay É. Scikit-learn: Machine Learning in Python.
\emph{Journal of Machine Learning Research.} 2021; 12:2825-2830. \\
{[}19{]} & Lavanya K, Rambabu P, Suresh GV, Bhandari R. Gene expression
data classification with robust sparse logistic regression using fused
regularisation. \emph{International Journal of Ad Hoc and Ubiquitous
Computing,} 2023; 42(4):281-291. \\
{[}20{]} & Dexter G, Khanna R, Raheel J, Drineas P.~Feature Space
Sketching for Logistic Regression. \emph{arXiv preprint,} 2023;
arXiv:2303.14284. \\
\end{longtable}
}

\end{document}